\documentstyle{article}
\begin{document}
\title{The Cavendish Experiment in General Relativity\thanks{Contribution
to Festschrift volume for Engelbert Schucking, to be published by Springer
Verlag}}
\author{Dieter Brill\\  {\small Department of Physics}\\
{\small University of Maryland}\\
\small College Park MD 20742, USA}

\date{}

\maketitle
\begin{abstract}
Solutions of Einstein's equations are discussed in which the ``gravitational
force" is balanced by an electrical force, and which can serve as models for
the Cavendish experiment.
\end{abstract}
\section{Introduction}
One of many useful lessons one can learn from Engelbert is the
appreciation of simple situations and examples that nonetheless can
teach us valuable physics. For me, such an Engelbert lesson was an
introduction to the Bertotti-Robinson universe (which, as Engelbert
added with characteristic precision that extends also to the history
of physics, was first discovered by Levi-Civita), and its relation to
extremal solutions \cite{Eng}. Below is a bit of physics that we can
learn from extremal solutions to Einstein's equations.

In General Relativity there is a well-defined sense in which the
equations of motion for particles follow from the field equations.
This is well known but not easily checked out,\footnote{I really mean
{\em nachvollziehbar}, a fashionable German word that seems to have no
good English equivalent.} for the necessary manipulations are rather
formidable.  It has been remarked \cite{Al} that when predictions of
General Relativity are based on particle equations of motion they
appear to lack the transparency and cogency that we appreciate in
Newtonian physics and in some alternative theories; that even the
outcome of the Cavendish experiment has not been derived in a way that
is both simple and rigorous; and that, at least in the case of
two-dimensional (``planar") translational symmetry there exist
``anti-Cavendish" solutions of Einstein's field equations, describing
slabs that do not attract each other. (In these solutions there are no
other interactions than gravity between the slabs, but the
stress-energy of the matter is ``exotic.")

In the present contribution we examine a question suggested by these
consid\-erations\footnote{I thank Prof. C. Alley (University of
Maryland) for numerous discussions which called attention to the
status of the Cavendish experiment {\sl vis a vis} General Relativity,
and in which he supplied the experimental ideas mentioned in passing
below.}, namely whether there are simple models in General Relativity
that are relevant to the Cavendish experiment. We will construct one
such model that is easily analyzed and whose predictions agree with
the expected experimental outcome. These models are not confined to
the plane symmetric case for which they were first discussed, and they
have no connection with the ``exotic" slab solutions. Nevertheless we
begin with a few elementary remarks about the special status of planar
symmetry in General Relativity as compared to other field theories.

\section{Planar symmetry}

In electrostatics, problems with planar symmetry (such as two parallel
charged plates) are among the simplest to treat.  The translational
and rotational symmetry of the physical setup prevents dependence of
physical quantities on the transverse ($y$, $z$) directions. The
problem therefore becomes one-dimensional. Physically one cannot, of
course, realize strict translational symmetry, because the system's
total charge and mass would be infinite. However, one can approximate
the one-dimensional situation by systems whose properties are
independent of $y$ and $z$ out to some large distance $D$, when one
considers only longitudinal distances $x$ small compared to $D$. In
the limit $D \rightarrow \infty$ the one-dimensional approximation
becomes arbitrarily accurate, and reasonable physical quantities have
finite limits. These include the electric field, the force per area,
and the acceleration of the plates.  (However, when the total charge
on the plates is non-zero, the electrostatic potential does not have a
finite limit, if it is normalized to zero at infinity.)

Another simple feature of translationally symmetric electrostatics is
the uniqueness of the relative acceleration between the plates. (The
electric field is unique up to an additive constant.) If one has {\it
any} solution with the appropriate symmetry, it is the {\it correct}
solution.\footnote{This is true provided the plates are indeed static.
In a typical experiment one balances the electric force between plates
by an elastic force, and measures how much elastic force is needed to
keep the plates static. If plates of finite size and non-negligible
charge were allowed to accelerate, they would of course radiate. The
radiation reaction would affect the plates' net acceleration, and this
would depend on radiation conditions imposed at infinity.} This
simplicity makes the (approximately) parallel plate geometry so useful
in both pedagogy and practice.

In Newtonian gravitation the situation is essentially identical;
everything one knows about electrostatics can be taken over (with the
appropriate sign of the force), except that there is no arbitrary
charge to mass ratio --- the (strong) principle of equivalence fixes
the ratio of gravitational to inertial mass to be a positive constant.
One might expect that the simplicity of the parallel plate geometry
will also carry over in General Relativity.

There are of course important differences between these theories,
which can destroy the analogy. One relevant difference that is usually
cited is the role of the potential. In electrostatics and in Newtonian
gravity the potential has no direct physical meaning, separate from
the fields.  In general relativity the analogous quantity is the
metric, and it measures directly the physically meaningful space-time
distances. When the size $D$ of the system increases (with constant
mass density), the Newtonian potential, normalized to zero at
infinity, typically diverges. This is not a serious problem in
electrostatics or in Newtonian gravity, because another normalization
can be chosen with impunity. However, a diverging metric offers more
serious problems, and is certainly not allowed in an asymptotically
flat spacetime. On the other hand, it is not obvious that this
divergence cannot be undone in the limit by suitable gauge changes;
and in any case one can take the view that if translationally
symmetric solutions exist, they should (approximately) describe
physically realistic parallel plates, since the General Relativity
solutions for finite plates presumably exist.

Static solutions with plane symmetry have in fact been studied in
general relativity, for example by Taub \cite{T} for matter with a
fluid equation of state. Unfortunately they do not readily lend
themselves to physical interpretation of the type sought here. (One
can however show on the basis of this work that, as expected, no
solution exists with vanishing pressure and positive mass density.)
Also, the relation of these solutions to any description of finite
parallel plates with proper (asymptotically flat) behavior at infinity
is not transparent.

\section{Exact, static solutions}

Static solutions would not seem to present a very versatile arena for
exploring the features of the gravitational interaction. They do
however correspond to a possible physical arrangement that would
reasonably be used in a sensitive experiment to measure the strength
of the gravitational interaction ($G$).  In the usual Cavendish
experiment,\footnote{Prof. Alley (private communication) points out
that this could be modified to realize the plane-symmetric geometry by
replacing the usual masses with parallel plates, one of which is
suspended (for example, by means of the traditional torsion fiber) so
that the total force on it can be monitored. This geometry has several
advantages, for example that the distance between the plates does not
have to be known with great accuracy, and that many of the devices
used in a parallel plate electrostatic measurement to increase the
accuracy, such as ``guard rings" to make the field more uniform, could
be adapted to the gravitational experiment. However, I do not know of
any attempt to obtain a more accurate measurement of $G$ in this way.}
even if initially the proof mass is in free fall, the long-time
behavior is typically governed by an interplay of gravitational
interaction and torsion fiber reaction. The initial acceleration can
generally not be measured as accurately as the final displacement,
which one can model as masses with constant separation --- in other
words, a static situation.

The gravitational interaction is then measured by the force necessary
to keep the masses apart, and the basic nature of this force is
electrical.  (One could replace the force of the torsion fiber by the
explicitly electrical force obtained, for example in the parallel
plate version of the experiment, by putting equal charges on the
plates.) We model this force by assuming that each volume also carries
a net charge, proportional to the mass of that volume, and all of the
same sign. For a suitable choice of the constant charge/mass ratio,
the attractive gravitational and repulsive electrical forces will then
balance in the Newtonian description.

How do we describe this situation in General Relativity? Because mass
and charge is present we must solve the Einstein and the Maxwell
equations, as well as the equations of motion of the matter. The
source in the Einstein equations is the stress-energy of the electric
field and that of the matter; the source in the Maxwell equations is
the charge density of the matter; and these equations imply the matter
equation of motion, at least for the simplest kind of matter,
``charged dust".\footnote{The static sources in these solutions are
unstressed, due to the detailed balance between electric and
gravitational forces. So we can imagine that these are elastic bodies
rather than dust, but with vanishing stress and strain. The
stress-energy tensor is then the same as that of dust, and the
solution still applies.  It is clear that in this model the
stress-energy tensor satisfies all energy conditions one might
reasonably want to impose.}  If it is indeed possible to balance the
forces in detail --- an expectation discouraged by the nonlinear
nature of Einstein gravitation, but encouraged by the absence of
interaction energies in the the corresponding Newtonian situation ---
then there should be a static solution. It is a remarkable theorem
\cite{CR} that such solutions not only exist, but that the fields have
a unique form under these conditions, the Majumdar-Papapetrou form
\cite{MP} that is well-known when gravity is generated not by matter
but only by charged black holes (and the stress-energy of their
electric field).  In the latter case the geometry and field can
represent any static arrangement of a finite number of black holes
with an {\em extremal} charge.

The Majumdar-Papapetrou ansatz for the metric and field can be written as
\begin{equation}
ds^2 = -V^{-2}dt^2 + V^2\left(dx^2+dy^2+dz^2\right) 
\qquad A^{\mu}=V \delta^{\mu}_t
\end{equation}
By explicit computation\footnote{I am grateful to the group of Prof. C. 
Alley for providing me with many of the
results cited below, as obtained by their computer calculations. The
conclusions drawn from these calculations are my own and have not been
fully discussed with Prof.~Alley.}
of the Einstein tensor $G_{\alpha\beta}$
and the Maxwell stress-energy tensor $T^{\rm EM}_{\alpha\beta}$
one finds agreement of most of the components of the two, for
example
\begin{eqnarray*}
G_{xx} &=& V^{-2}\left(-V_x^2+V_y^2+V_z^2\right) = T^{\rm EM}_{xx} \\
G_{xy} &=& -2V_xV_y =T^{\rm EM}_{xy} \\ {\rm etc.}&&
\end{eqnarray*}
The exception is $(\alpha, \beta) = (t,t)$. Similarly one finds that the 
$A^\mu$ of Eq (1.1) satisfy most of the components of the vacuum 
Maxwell equations,
$${F_\alpha^\beta}_{;\beta} = 0 \quad \hbox{except for}\quad \alpha=t.$$
This structure of the field equations is appropriate for a static dust source, 
since that type of source contributes only
to the components that are excepted above.  For these one 
has the condition (in units where the gravitational and electromagnetic
coupling is unity)
\begin{equation}
\begin{array}{rcl}
G_{tt} - T^{\rm EM}_{tt} &=& -2V^{-5}\nabla^2V = T^{\rm matter}_{tt}\\
{F_t^{\beta}}_{;\beta} &=& -V^{-4}\nabla^2V = J^{\rm matter}_t.
\end{array}
\end{equation}
Here the Laplacian $\nabla^2$ is to be evaluated in the flat 
three-dimensional background metric $dx^2 + dy^2 + dz^2$.

All the equations are satisfied if charged dust can supply both sources 
in the equations (1.2). For this type of matter, with mass density $\rho$ 
and charge density $\sigma$, we have
\begin{eqnarray*}
T^{\rm matter}_{\alpha\beta}&=& \rho u_\alpha u_\beta \\
J^{\rm matter}_\alpha &=& \sigma u_\alpha.
\end{eqnarray*}
>From the metric (1.1) we find that the unit four-velocity for the static matter
has the form $u_\alpha = V^{-1} \delta^t_\alpha$. Thus we see that equations 
(1.2) are satisfied if we choose
\begin{equation}
\rho =\sigma = V^{-3}\nabla^2V.
\end{equation}

The equation relating the ``potential" $V$ and the source $\rho$ is very
similar to the Newtonian equation; in the vacuum region they are identical. 
One way to make a correspondence between the two is the following:
Given any Newtonian potential $V_N$ (vanishing at infinity)
and source $\rho_N$ one finds a solution of equation (1.3) by
$$V = 1 +V_N \qquad \rho = \rho_N/V^3.$$
Thus $\rho$ has the same support as $\rho_N$, and the two differ only
slightly if the gravitational fields are weak, $|V_N| \ll 1$.

This solution to the equations of general relativity has all the
physically reasonable properties one expects; but could there be
other solutions to the same problem with different properties? Suppose
any static solution to the Einstein-Maxwell equations is given. For
simplicity, confine attention to the region outside the matter.
Let the electrostatic potential $A^t$ of the solution be $V$. Let
$g_{ij}$ be the spacelike metric on the three-dimensional hypersurfaces 
that are orthogonal to the timelike Killing vector. One can
then show that when one modifies the metric by a conformal factor $V^{-2}$, 
its Ricci tensor vanishes, $R_{ij}[V^{-2}g_{ij}] = 0$. The three-dimensional 
modified metric must therefore be flat,
and hence the original metric and field must have the form of Eq (1.1).
In this sense, then, the solutions given above are unique.

\section{Test particle motion}

It is not necessary to verify separately that the equations of motion 
for the matter are satisfied by the solution given by equations (1.1, 1.3),
because the matter motion is a consequence of the field equations.
However, as a further check that this solution is reasonable, we derive
the equation of motion for a test particle in the fields of this solution.

As in Newtonian physics, the general relativistic motion of test particles
in the general metric (1.1) (even with V harmonic) is not integrable, but
there is always an energy integral. The energy integral is enough to find 
the motion if we know that it is confined to one coordinate line. This
is the case for example when there is planar ($x,\,y$) or axial (about
the $z$-axis) symmetry. We therefore confine attention to these cases
where the energy integral yields the essential information about the motion.

In the case of uncharged particles we can apply the usual theorems about
geodesics, that any Killing vector like $\partial/\partial t$ yields a
conservation law of the corresponding covariant component of the 4-velocity $u$,
$u_t = -E$. From the metric (1.1) we therefore have, with $\tau=$ proper time
$$ E = -g_{tt}{dt\over d\tau}= {1\over V^2}{dt\over d\tau} .$$
We also know that setting the length of $u$ to unity is always an integral 
of the equation of motion,
\begin{equation}
u\cdot u =-1 = -{1\over V^2}\left({dt\over d\tau}\right)^2 + 
V^2\left({dx\over d\tau}\right)^2  .
\end{equation}
Elimination of $dt/d\tau$ yields
\begin{equation}
\left({dx\over d\tau}\right)^2 + {1\over V^2} = E^2.
\end{equation}
So for geodesic motion the quantity $1/V^2$ acts as an effective potential,
and the particle will be
deflected from the Killing orbit $(x,\,y,\,z)=$ const by an amount
proportional to $\nabla V$, as in the Newtonian description. 

For weak fields we have $V \approx 1 - \Phi$ where $\Phi$ is the Newtonian
potential (for a spherically symmetric mass $M$, $\Phi = -M/r$), hence
$${\textstyle{1\over 2}}\left({dx\over d\tau}\right)^2 + 
\Phi \approx{\textstyle{1\over 2}} (E^2-1),$$
which is (essentially) the usual energy integral for potential motion,
yielding the usual motion corresponding to attractive gravity.

When the particle is charged (with charge/mass $q$) the corresponding 
conservation laws are derived from the variational principle
$$\delta \int (u \cdot u + 2q u \cdot A) d\tau = 0.$$
If there is a Killing vector that leaves the metric and the potential $A$
invariant,  there is a conserved quantity (Noether's theorem); for the 
Killing vector $\partial/\partial t$ the conserved quantity is the
momentum conjugate to t,        $-E = u_t + qA_t$, or
$$E = {1\over V^2}{dt\over d\tau} -{q\over V}.$$
We substitute this in (1.4), eliminate $dt/d\tau$, and find 
$$\left({dx\over d\tau}\right)^2 + {1-q^2\over V^2}+{2Eq\over V}=E^2.$$
We see that the attractive gravitational potential of the comparable
equation (1.5) is reduced, and becomes repulsive for $q^2 > 1$. There
is also a contribution to the potential that is proportional to $E$.

In the special (``extremal"\footnote{In this
context the term ``extremal" is somewhat misleading: it is the maximum charge
that a black hole of the given mass could have, but is not a large charge
at all for that mass of ordinary matter to carry.}) case $q^2 = 1$ 
and $E = 0$, an initially static test particle remains at rest at any position 
(as does the matter that produces these fields). This is the case where 
attractive gravity and repulsive electrostatics are in perfect balance. 
If the particle is initially not quite at rest, then $Eq$ must be slightly 
negative (since $V > 0$), hence the term $2Eq/V$ represents an attractive 
potential. It may be interpreted as a response of the particle's increased
``relativistic mass'' to gravity, with no compensating increase in the
particle's charge. 

\section{Conclusion}

We have exhibited the unique static solutions to the
Einstein-Maxwell-matter field equations that represent an arbitrary
distribution of extremally charged matter in the form of dust. In
particular, these solutions can be a good approximation to the
geometry of a Cavendish experiment.  Because all the charges have the
same sign, the electric interaction is repulsive between two volumes
of matter.  The constancy in time of the physical distance between the
masses implies, in ordinary language, a balancing attractive
gravitational interaction.  In this sense we have shown that in
general relativity, as in Newtonian gravity, the gravitational
interaction between the bodies is nonzero and attractive. Because the
solution is valid only when the charge has the extremal value, such a
balancing Cavendish experiment could be used to find the extremal
charge value for a given mass.\footnote{Results from an electrically
balancing Cavendish experiment have recently been reported \cite{FA};
however, that experiment did not measure the extremal charge value
because (for good reasons) the attracting ``large mass'' was not the
same as the repelling electrode.} Measurement of this extremal charge
to mass ratio is equivalent to measuring the gravitational constant
$G$.

\end{document}